\documentclass[%
reprint,
 amsmath,amssymb,superscriptaddress,
longbibliography,
prl
]{revtex4-1}
\usepackage{xparse}

\usepackage{appendix}
\usepackage{graphicx}
				\graphicspath{ {./images/} } 
\usepackage{dcolumn}
\usepackage{bm}
\usepackage{hyperref}

\usepackage{slashed}
\renewcommand{\vec}[1]{\mathbf{#1}}

\usepackage{bbold}
\usepackage{xcolor} 

\begin{document}
\title{Strained Bilayer Graphene, Emergent Energy Scales, and Moir{\'e} Gravity}

	\author{Alireza Parhizkar}
\affiliation{Joint Quantum Institute and
	Condensed Matter Theory Center, University of Maryland, College Park, MD 20742, USA}
\author{Victor Galitski}
\affiliation{Joint Quantum Institute and
	Condensed Matter Theory Center, University of Maryland, College Park, MD 20742, USA}

\date{
    \today
}

\begin{abstract}
Twisted bilayer graphene is a rich condensed matter system, which allows one to tune energy scales and electronic correlations.  The  low-energy physics of the resulting  moir{\'e} structure can be mathematically described in terms of a diffeomorphism in a continuum formulation. We stress that twisting is just one example of moir{\'e} diffeomorphisms. Another particularly simple and experimentally relevant transformation is a homogeneous isomorphic strain of one of the layers, which gives rise to a nearly identical moir{\'e} pattern (rotated by $90^\circ $ relative to the twisted structure) and potentially flat bands. We further observe that low-energy physics of the strained bilayer graphene takes the form of a theory of fermions tunneling between two curved space-times. Conformal transformation of the metrics results in  emergent  ``moir{\'e} energy scales,'' which can be tuned to be much lower than those in the native theory. This observation generalizes to an arbitrary space-time dimension with or without an underlying lattice or periodicity and suggests a family of toy models of ``moir{\'e} gravity'' with low emergent energy scales. Motivated by these analogies, we present an explicit toy construction of moir{\'e} gravity, where the effective cosmological constant can be made arbitrarily small. We speculate about possible relevance of this scenario to the fundamental vacuum catastrophe in cosmology.

\end{abstract}
\date{\today}

\maketitle


When two lattices overlap, they give rise to a moir\'{e} pattern as their emergent superlattice structure. The physical properties of such moir\'{e} superstructures have  been extensively studied in the context of bilayer graphene~\cite{Exp2018correlated,Exp2018unconventional,GrapheneWithTwist,GrapheneWithTwist2020,ReviewOpportunities,ElectronicSpectrum,BandStructure,PiezoStrain}. The low-energy physics of graphene bilayers can be conveniently studied within a continuum model, where a general deformation is described by the two-dimensional diffeomorphism $\vec{x} \rightarrow \vec{x} + \vec{\bm \xi}(\vec{x})$ where points at $\vec{x}$ are translated by $\vec{\bm \xi}$, which in general can be an arbitrary function of the position vector $\vec{x} \equiv (x,y)$. Starting with two layers with coinciding sites, deforming one of the layers by the flow $\vec{\bm \xi}(\vec{x})$ yields a general  bilayer superstructure. Specifically for the  twisted bilayer graphene, the twist flow for small twist angles $\theta$ is given by $\vec{\bm \xi}_t \equiv \theta \hat{z} \times \vec{x}$. 

\begin{figure}
\includegraphics[width=\linewidth]{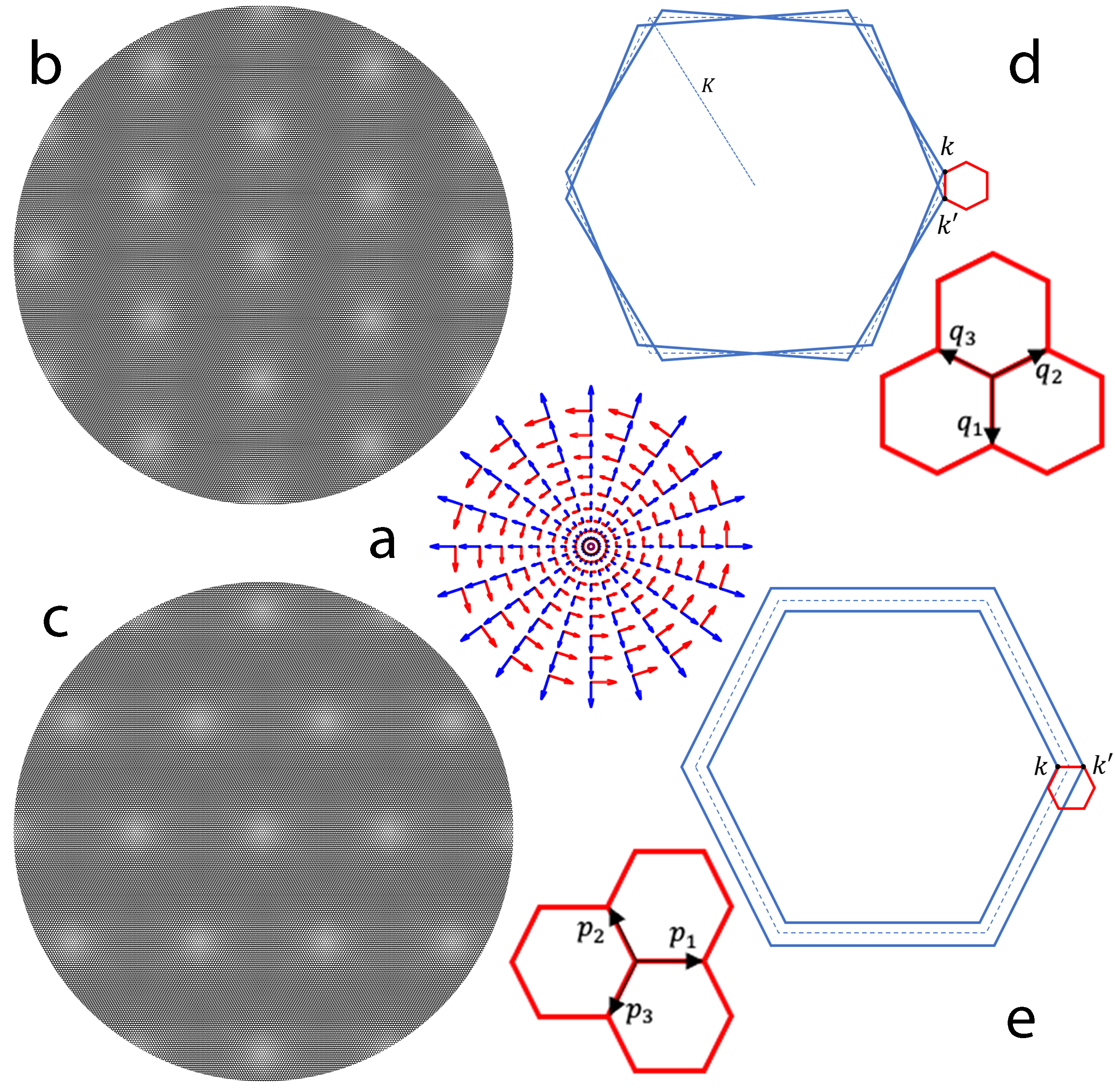}
\centering
\caption{(a) Generators of rotation and expansion drawn simultaneously: Blue arrows demonstrate the vector field $\vec{\xi}_s$ while red arrows demonstrate that of $\vec{\xi}_t$. (b) and (c) are twisted and stretched bilayer graphene respectively. The parameters of both transformations are set to $\theta \approx 0.0192$. Moir\'{e} patterns of these bilayers are exactly the same but $90^\circ$ rotated. Hexagonals (unit cells) of undeformed layers have their largest diameter along the vertical axis $y$. (d) and (e) schematically show Brillouin zones of twisted and stretched bilayer respectively (blue hexagonals); both give rise to their corresponding moir\'{e} reciprocal lattices (red hexagonals)}
\label{fig:Together}
\end{figure}

Much similar to this flow, but perpendicular to it  is the flow due to a biaxial strain or uniform expansion of the layers. It is described by $\vec{\bm \xi}_s \equiv \theta \vec{x}$, where we use the same notation $\theta$ for the expansion parameter. The strained and twisted vector fields are related to each other by a $90^\circ $ rotation, $\vec{\bm \xi}_t \cdot \vec{\bm \xi}_s = 0$,  as shown in Fig \ref{fig:Together}a. Furthermore, since the transformations are similar but orthogonal to each other, the corresponding emergent moir\'{e} patterns are also $90^\circ $ rotated versions of each other. This is also shown in Fig. \ref{fig:Together} where we have deformed one layer by $+\frac{\theta}{2}$ and the other layer by $-\frac{\theta}{2}$.  Note that the combinatory effect of twist and different types of strain have been investigated both experimentally and theoretically in Refs.~[\onlinecite{ExpStrainFields,ExpStrainCurvature,Designing,StrainedBilayer,TvsStraintronics,StrainProperties}]. 

The main goal of the present Letter is to generalize this construction to  a broad class of systems, by pointing out that moir\'{e} length and energy scales generally emerge in continuum theories, where two smooth manifolds  of arbitrary  dimension overlap or are coupled together, and where the metric of one of the manifolds is a scaling diffeomorphism of the other. No underlying lattice structure, nor quantum mechanics are necessary for this purely geometric phenomenon to occur. However, since the appearance of large moir\'{e} patterns and small moir\'{e} bands in slightly strained bilayer graphene is worthy of explicit emphasis, and given the direct relevance to experiment, we first review the specific physics of uniformly strained bilayer graphene, with the main intention of following the geometric emergence of the small energy scales. Most results are straightforwardly transplanted from the case of twisted bilayer graphene, and so are discussed/reviewed in parallel.

\begin{figure}
\includegraphics[width=\linewidth]{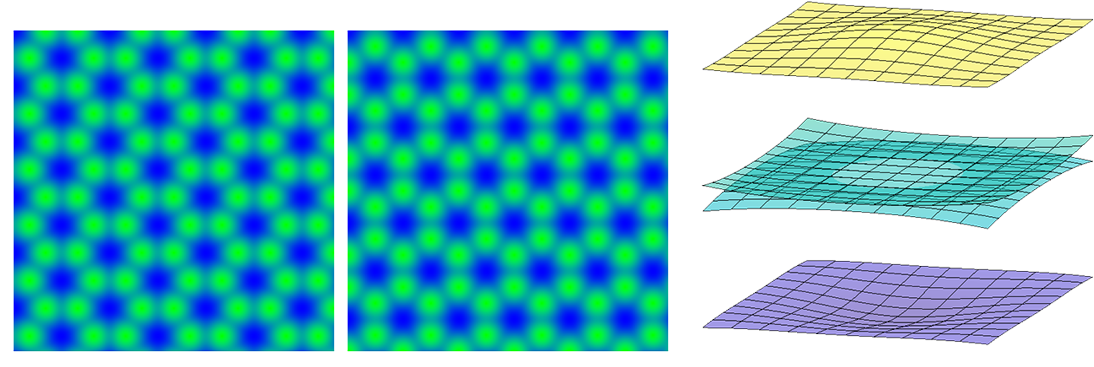}
\centering
\caption{From left to right: Tunneling matrix for twisted bilayer - Tunneling matrix for stretched bilayer - Vanishing of the renormalized Fermi velocity at K valley for magic scale $\theta \approx 0.0192$. Green designates regions where AB tunnelings are dominant and blue shows that of AA tunnelings.}
\label{fig:TunFlat}
\end{figure}

The continuum model for both twisted and strained bilayer  is given by the following Hamiltonian~\cite{balents} 
\begin{align}
		H_{t,s} = \int d^2x \bigg[ &\psi_+^\dagger h_{t,s}^{+\theta/2} \psi_+ + \psi_-^\dagger h_{t,s}^{-\theta/2} \psi_-  \nonumber \\
		&+ \psi_-^\dagger T_{t,s} (\vec{\xi_{t,s} }) \psi_+ + h.c. \bigg] \, ,
		\label{Hamiltonian}
\end{align}
where $\psi_\pm$ are fermionic operators, $L=\pm$ indexes the upper/lower layers,  and $h^{\pm\theta/2}_{t,s}=-iv_F \vec{\bm \sigma}_{t,s} \cdot \vec{\bm \nabla}$ is the single-particle Hamiltonian in layer $L=\pm$ deformed by $\pm\frac{\theta}{2}$ under the flow $\vec{\bm \xi}_t$ or $\vec{\bm \xi}_{s}$. Also $v_F$ is the Fermi velocity and $\vec{\bm \sigma}_{t,s}$ is $\vec{\bm \sigma}=(\sigma^x,\sigma^y)$ transformed accordingly. Note that all the fields in the above equation depend on position. The inter-layer tunneling matrix $T$ has two parts, a diagonal part proportional to $\sigma^0$ describing intra-sublattice (AA) tunneling and an off-diagonal part describing inter-sublattice (AB) tunneling. This results into two types of eigenvalues for the tunneling matrix, which overcome each other periodically over the moir\'{e} pattern. Therefore we expect the eigenvalues of the tunneling matrix for twisted and stretched bilayer to schematically follow Fig. \ref{fig:TunFlat}.

The tunneling matrix $T_t$ for twisted bilayer graphene in real space is given by $T_t (\vec{x}) = \sum_j e^{-i \vec{q}_j \cdot \vec{x}} T^t_j$ with $\vec{q}_1 = k_\theta (0,-1)$, $\vec{q}_{2,3}=k_\theta (\pm \sqrt{3}/2,1/2)$ and  \cite{balents,Origin} 
\begin{equation}
	T^t_j = u 
	\begin{pmatrix}
		1 & 0 \\
		0 & 1
	\end{pmatrix}	
	 + w 
	\begin{pmatrix}
		0 & e^{-i\frac{2\pi}{3}(j-1)} \\
		e^{i\frac{2\pi}{3} (j-1)} & 0
	\end{pmatrix} \, ,
\end{equation}
where $u$ and $w$ are respectively AA and AB inter-layer coupling parameters and $k_\theta \equiv K\theta =\frac{4\pi}{3\sqrt3 a} \theta$. Since $T_s$ is a $90^\circ$ rotated version of $T_t$ then one expects $T_s (\vec{x}) = \sum_j e^{-i \vec{p}_j \cdot \vec{x}} T^s_j$ with $\vec{p}_1=k_\theta (1,0)$ and $\vec{p}_{2,3}= k_\theta (-1/2,\pm \sqrt{3}/2)$. 

The Brillouin zone of a single layer of graphene is hexagonal with two valleys K and K$'$ where Dirac cones reside. A transformation of a single layer in real space, transforms the reciprocal lattice accordingly. In the twisted bilayer, when one layer is rotated by $+\frac{\theta}{2}$ and the other by $-\frac{\theta}{2}$, the corresponding reciprocal lattices of the two single layers also rotate by the same angles. This is due to the fact that rotation applies to all vectors; therefore it will also take place in the momentum space where $\vec{k} \equiv (k_x,k_y)$ has rotated in the same way as $\vec{x}$. If we concentrate only on one valley, say K, then we see that rotation separates the K valleys of the two layers by $k_\theta$ as shown in Fig. \ref{fig:Together}d.

The same happens for the uniformly strained bilayer. The single layer, stretched by $+\frac{\theta}{2}$ in real space, experiences a shrinkage in momentum space by the same factor, and the inverse effect happens to the other layer. Since, like the above case of twisted bilayer, we have respected all the symmetries of graphene through this deformation, the hexagonal structure is preserved. Therefore, a moir\'{e} reciprocal pattern emerges similar to twisted bilayer but rotated by $90$ degrees, see Fig~\ref{fig:Together}e.

By looking at the moir\'{e} reciprocal lattices of the both bilayers, figures \ref{fig:Together}d and \ref{fig:Together}e, we see that there are three different paths an electron can take while tunneling between layers from k to k$'$. The three-momentum vectors $\{\vec{p}_1,\vec{p}_2,\vec{p}_3\}$ in strained bilayer are the $90^\circ$ counterclockwise rotated version of $\{\vec{q}_1,\vec{q}_2,\vec{q}_3\}$ in twisted bilayer. An electron at the end of $\vec{p}_1$ compared to an electron at the end of $\vec{q}_1$ is rotated by $90^\circ$ or by $e^{i\frac{\pi}{4}\sigma^z}$. If $T_s$ is indeed related to $T_t$ by a rotation then we should be able to write
\begin{equation}
	T_s(\vec{\rm \xi}_s) = e^{-i\frac{\Omega}{2}\sigma^z} T_t \left( R(\Omega) \vec{\bm \xi}_t \right) e^{i\frac{\Omega}{2}\sigma^z} \, , \ \ \Omega=\frac{\pi}{2} \, , \label{TT}
\end{equation}
where both spinor degrees of freedom of $T_t$ and its argument are rotated by $90^\circ$. Then one can numerically \cite{MacDonald} and analytically \cite{Origin} (for the chiral limit where $u=0$) show that there are magic scales $\theta$ where the band structure of the stretched bilayer develops a flat band; see Fig.~\ref{fig:TunFlat}. Note however that if the tunneling matrix does not satisfy condition~(\ref{TT}), flat bands would not necessarily appear. 

Now, we rewrite the Hamiltonian for a bilayer deformed by an arbitrary flow ${\bm \xi}_{\pm}({\bf x})$  in the following geometric form
\begin{align}
		H = \int d^2x \bigg[ &\psi_+^\dagger e^\mu_{l+} \sigma^l D^+_\mu \psi_+ + \psi_-^\dagger e^\mu_{l-} \sigma^l D^-_\mu \psi_-  \nonumber \\
		&+ \psi_-^\dagger T_\xi \left( \vec{\xi}_+(\vec{x}) - \vec{\xi}_-(\vec{x}) \right) \psi_+ + h.c. \bigg] \, ,
		\label{CurvedHamiltonian}
\end{align}
where $D^\pm_\mu =\partial_\mu + \mathcal{A}^\pm_l e^l_{\mu\pm}$ with $e^\mu_{l\pm} = \delta^\mu_l + \frac{\partial \xi_\pm^\mu}{\partial x^l}$ being the vielbeins of the two-dimensional space of each single layer deformed by $\vec{\xi}_\pm$ and $\mathcal{A}^\pm_l$ gauge fields induced by the flow (e.g. non-uniform strain) \cite{ElectronicProperties,Strain} in each layer.
 Explicitly, for a single layer, $\mathcal{A}_l$ is defined as
\begin{equation}
	\mathcal{A}_l = \frac{\gamma}{a}
	\begin{pmatrix}
		\partial_x \xi_x - \partial_y \xi_y \\
		\partial_x \xi_y + \partial_y \xi_x
	\end{pmatrix}
	, \ \ \gamma \equiv \frac{\partial \ln t}{\partial \ln a}
\end{equation}
where $t$ is the hopping strength.
So far the geometry of each single layer has been considered to be flat. However, due to the robustness of the Dirac points we can still include slight deviations from the flatness to our considerations as well.
\begin{figure}
\includegraphics[width=\linewidth]{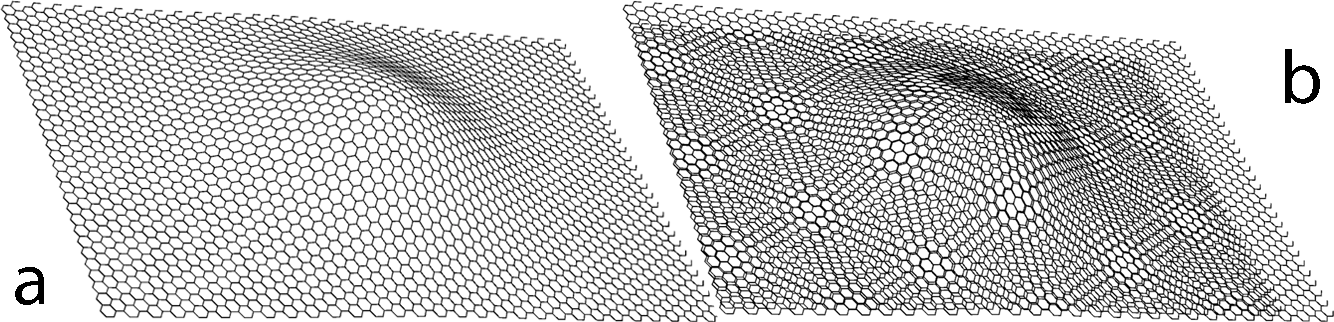}
\centering
\caption{(a) A single curved layer of graphene. (b) A bilayer constructed from two equally curved layers with unequal graphene charts. One chart is slightly shrunken relative to the other which has resulted in the familiar moir\'e patterns.}
\label{fig:CBilayer}
\end{figure}
We can imagine a curved layer of graphene as a curved surface covered by a hexagonal chart of graphene. The diffeomorphism flow, $\bm\xi$, transforms the chart to another leaving the surface untouched. To consider a curved bilayer, therefore, we use two relatively transformed curved vielbeins $e^\mu_{l\pm} = e^\mu_l + \frac{\partial \xi_\pm^\mu}{\partial x^l}$ with the original curved vielbein $e_l^\mu$ yielding the induced metric $h^{\mu\nu}$ on the surface through $h^{\mu\nu}= e^\mu_m e^\nu_n \delta^{mn}$ where $\delta^{mn}$ is the metric of the flat space. If we describe the two dimensional curved surface as a membrane embedded in a three dimensional flat manifold, Fig. \ref{fig:CBilayer}, whose points satisfy the relation $z=h(x,y)$ with $z$ being the extra dimension, then the induced metric $h_{\mu\nu}$ on this surface is given by,
\begin{align}
    h_{\mu\nu} dx^\mu dx^\nu = &\left(1+(\partial_x h)^2\right) dx^2 \nonumber \\
    &+ 2 \partial_x h \partial_y h dx dy + \left(1+ (\partial_y h)^2\right) dy^2 \, .
\end{align}
Because of the curvature, $\pi$-orbitals which where aligned before will now make a locally varying angle with each other and consequently result in a locally varying hopping parameter which ultimately gives rise to an additional gauge field $\mathcal{A}^c_l$ appearing in the same way as $\mathcal{A}_l$ inside the covariant derivative. For small $h(x,y)$ it is given by
\begin{equation}
    \mathcal{A}^c_l = \epsilon_{\pi\pi}\frac{3a^2}{8}
	\begin{pmatrix}
		(\partial_y^2 h)^2 - (\partial_x^2 h)^2 \\
		2\partial_x\partial_y h (\partial_x^2 h + \partial_y^2 h)
	\end{pmatrix} \, ,
\end{equation}
with $\epsilon_{\pi\pi}=2.89\text{eV}$ for graphene.\cite{MembraneGraphene,ElectronicProperties} Note that one can also attribute unequal curvatures to the layers by considering them having unequal induced metrics. In that case the difference in curvature itself will automatically introduce a moir\'e pattern.

Parallel transportation of fermions on a curved spacetime is determined by the covariant Dirac operator $ \slashed{D} \equiv e^\mu_k \sigma^k \left( \partial_\mu + A_\mu^{mn} [\sigma_m,\sigma_n] \right)$. However, using the algebra of Pauli matrices and disregarding time components we can write this Dirac operator as $\slashed{D} = e^\mu_k \sigma^k \left(\partial_\mu + \mathcal{A}_l e^l_\mu \right)$ with $\mathcal{A}_l$ defined as $\mathcal{A}^l \equiv e^\mu_m A_\mu^{[ml]}$. Therefore, written in the above form \eqref{CurvedHamiltonian}, the bilayer problem resembles a fermionic field theory where the electron is allowed to tunnel between two different ``universes'' with their designated geometries given by the metrics $g^{\mu\nu}_\pm = e^\mu_{m\pm}e^\nu_{n\pm}\eta^{mn}$. Here $\eta^{mn}$ is the flat metric. The interlayer tunneling plays the role of a ``wormhole'' process in this formulation.\footnote{Of course, by this we do not mean the formation of an Einstein-Rosen bridge. Metric is not a dynamical field in \eqref{CurvedHamiltonian}, instead one can imagine a potential barrier between the layers that fermions are allowed to pass through by the tunneling process.}

This perspective allows us to identify minimal ``ingredients'' needed for   moir\'{e} physics, which we  define as the  emergence of new energy/length scales in two superimposed systems, much smaller/larger compared to the corresponding scales in the individual systems. We observe that no underlying lattice is necessary, and  two continuum models (even with a random structure) may give rise to similar phenomena. We further notice that these conclusions are independent of dimensionality. 
These general considerations motivate us to broaden the scope of physical models, where these scenarios can be explored, beyond canonical condensed matter systems. A class of models where the moir\'{e} scenario may be of potential interest is general relativity/cosmology, in particular the cosmological constant problem. Below we propose a toy model where superimposing ``universes'' with large individual cosmological constants gives rise to an arbitrarily small effective cosmological constant.

 The problem with the cosmological constant $\Lambda$ can be stated in multiple ways (for a review, see e.g. Refs.~[\onlinecite{CCBurgess}] and [\onlinecite{CCPadilla}]). One of which is to ask: How can one naturally get a small or zero value for the gravitating cosmological constant, while the scales of the theory are huge in comparison? All the fields in the Standard Model  contribute to the zero-point energy density. Because gravity couples to all forms of energy, the gravitational effects of the zero-point energy must in general be observable. But this is not consistent with observations, which show the energy density  much smaller than all other scales in the Standard Model. On the other hand, the history of the universe and the resulting cosmology can be very sensitive to a model chosen to describe  the vacuum energy. Trying to reconcile all these with observations leads to a fine tuning problem. In the light of such problems then, the possibility of an emergent small scale might be a question of interest. What follows is a simple model inspired by the moir\'{e} physics, which can investigate this possibility. Note that the toy model of ``moir\'{e} gravity'' below is not unique and other similar setups can be constructed with different mechanism of overlapping geometries.


The classical theory of general relativity in the presence of matter fields is given by the following action devised for the Neumann boundary conditions,
\begin{equation}
	S_c = \int d^4x \sqrt{-g}\left[ \frac{c^4}{8\pi G} \left( R - 2\Lambda \right)  + \mathcal{L}(\phi) \right] \, ,
\end{equation}
where $g$ is the determinant of the metric $g_{\mu\nu}$, $\mathcal{L}(\phi)$ is the Lagrangian of all other fields in the theory, $c=1$ is the speed of light and $G= M_P^{-2}$ is the constant of gravitation with $M_P$ being the Planck mass. The largest scales of the theory are set by $M_P$. In four dimensional spacetime, then, a dimensionless action is provided only if $\Lambda$ scales with mass squared which sets the conjectured value of $\pm M_P^2$ for the cosmological constant. A more careful QFT consideration of the zero-point energy also casts the same guess. It is important to note that although it is to some extent meaningful to decide about the magnitude of a variable such as $\Lambda$ by the scales of the theory, one cannot decide about the sign of that variable using only the scales. Nonetheless, $+M_P^2$ and $-M_P^2$ are way larger than what  the cosmological constant is observed to be.

The Planck mass also roughly represents the upper bound for energy scales of measurement, or inversely, how accurately we can measure length. As we approach the limit of $M_P^{-1}$ the theory is expected to break down and, quite similar to condensed matter systems, a micro-structure should reveal itself when the wavelengths of intended observations are no longer blind to the underlying structure. Therefore, if spacetime was a lattice, the distance between the sites would have roughly been $M_P^{-1}$. This incites the idea that a combination of two such structures with slightly different length scales, can give rise to another moir\' e  length scale much longer than the two.

Then let us consider two copies of the classical theory with different metrics, which one can picture as two copies of a universe with the combined following action,
\begin{align}
	S_g + S_h &= \int d^4x \sqrt{-g}\left[ \frac{c^4}{8\pi G} \left( R_g - 2\Lambda_g \right)  + \mathcal{L}_g(\phi) \right] \nonumber \\
		&+ \int d^4x \sqrt{-h}\left[ \frac{c^4}{8\pi G} \left( R_h - 2\Lambda_h \right)  + \mathcal{L}_h(\phi) \right] \, ,
\end{align}
where $g_{\mu\nu}$ and $h_{\mu\nu}$ are the metrics of the two universes. Here $S_g$ and $S_h$ are already coupled through matter fields $\phi$, but for the present let us forget about non-geometrical fields and instead introduce an inter-universe coupling via a purely geometrical coupling term
\begin{equation}
    S_{gh} = \frac{c^4}{2\pi G} \int d^4x   \sqrt{|g,h|} \, \bar{\Lambda} \, ,
\end{equation}
where $|g,h|$ is generically any term built out of metrics $h_{\mu\nu}$ and $g_{\mu\nu}$. In what follows we set $\frac{c^4}{8\pi G}=1$ and define $|g,h|$ to resemble $g$  and $h$ as below
\begin{align}
	|g,h| &\equiv \frac{1}{4!}\varepsilon^{\mu\nu\alpha\beta}\varepsilon^{\rho\sigma\lambda\gamma} g_{\mu\rho}g_{\nu\sigma}h_{\alpha\lambda}h_{\beta\gamma} \label{HalfBred} \, ,
\end{align}
where $\varepsilon^{\mu\nu\alpha\beta}$ is the Levi-Civita symbol. The crossbreed determinant \eqref{HalfBred} transforms akin to metric determinant  $g \equiv  \frac{1}{4!}\varepsilon^{\mu\nu\alpha\beta}\varepsilon^{\rho\sigma\lambda\gamma} g_{\mu\rho}g_{\nu\sigma}g_{\alpha\lambda}g_{\beta\gamma}$ under general coordinate transformations.  So, the two universes are coupled through a shared or mixed volume element.
Note that there is no ambiguity in lowering and raising indices. The metric $g_{\mu\nu}$ is responsible for that of objects belonging to $g$-universe and the same goes for $h_{\mu\nu}$ and $h$-universe. For conformally flat metrics, which suffice for our present purposes, $|g,h|$ becomes equivalent to $\sqrt[4]{gh}$. We have adopted the former here merely to point out a wider range of possibilities. However, looking back at the Hamiltonian \eqref{CurvedHamiltonian}, we can think of the spinor degrees of freedom as fermionic fields that each have absorbed half a corresponding vielbein determinant $\psi_\pm \equiv \sqrt{|e_\pm|} \tilde{\psi}_\pm$, while each veilbein determinant $|e_\pm|$ is itself equivalent to the square root of its corresponding metric determinant. In fact, the quantum variables, for example in path-integral formulation of fermions on curved background, must be of this type if the theory is to be diffeomorphism invariant. This establishes a common characteristic between inter-layer and inter-universe tunnelings. A conformal transformation of only one metric transforms the coupling term halfway through.

To obtain the equations of motion we vary the action $S=S_g + S_h + S_{gh}$ with respect to the inverse metrics $g^{\mu\nu}$ and $h^{\mu\nu}$. Since the action is symmetric under $g\leftrightarrow h$ we are only going to write one set of these equations in what follows unless required otherwise. Variation of $S_g$ and $S_h$ gives the usual Einstein tensor with their corresponding cosmological constants, and the variation of the coupling term is obtained by looking at \eqref{HalfBred} and noticing $\delta g^{\mu\nu} = -g^{\mu\alpha}g^{\nu\beta}\delta g_{\alpha\beta}$. The result is
\begin{align}
	&\sqrt{-g} \left( R^g_{\mu\nu} - \frac{1}{2}R^g g_{\mu\nu} + \Lambda_g g_{\mu\nu} \right) = \nonumber \\
	&\quad \frac{\bar{\Lambda}}{3!} |g,h|^{-\frac{1}{2}} \varepsilon^{\kappa\zeta\alpha\beta}\varepsilon^{\rho\sigma\lambda\gamma} g_{\mu\kappa}g_{\nu\rho}g_{\zeta\sigma}h_{\alpha\lambda}h_{\beta\gamma} \, \  \label{EEg}  \, .
\end{align}

Let us now settle to a class of solutions which enjoy a large amount of symmetry, by choosing the Friedmann–Lemaître–Robertson–Walker metric with a conformal time $t$,
\begin{align}
	ds^2_g = a_g^2(t) \eta_{\mu\nu}dx^\mu dx^\nu \equiv g_{\mu\nu} dx^\mu dx^\nu \, \  \\
	ds^2_h = a_h^2(t) \eta_{\mu\nu}dx^\mu dx^\nu \equiv h_{\mu\nu} dx^\mu dx^\nu \, ,
\end{align}
Note that even though the coupling action $S_{gh}$ is invariant under coordinate transformations, choosing both metrics to have the above form is a kinetic restriction since, for example, there are no coordinate transformations that can generally transform both metrics to have the form $ds^2 =  -dt^2 + a^2(t)d\vec{x}^2$. But we deliberately restrict ourselves to this class of double-metrics since they present the simplest pathway for our model. So, we arrive at two sets of equations. From \eqref{EEg} for $\mu=\nu=0$ we have
\begin{align}
	3\dot{a}^2_g - \Lambda_g a_g^4  - \bar{\Lambda}a_g^2 a_h^2 =0
\end{align}
and for $\mu=\nu=\{1,2,3\}$,
\begin{align}
	\dot{a_g}^2-2\ddot{a}_ga_g +\Lambda_g a_g^4  + \bar{\Lambda} a_g^2 a_h^2 =0 \, .
\end{align}
The solutions are
\begin{align}
	a_g=\frac{A_g}{t + \tau_g} \mbox{ and }
	a_h=\frac{A_h}{t + \tau_h} \, ,
\end{align}
with $A_{h,g}$ and $\tau_{h,g}$ being the constants of integration. A solution to the above equations does not always exist, but requires the consistency relation between $\Lambda$s to be satisfied. If we let $\Lambda$s depend on time it is given by
\begin{align}
	\Lambda_g = \Lambda^g_\text{eff} - \frac{\Lambda^{g}_\text{eff}}{\Lambda^h_\text{eff}} \left(\frac{t + \tau_g}{t + \tau_h}\right)^2 \bar{\Lambda} \, ,
\end{align}
where $\Lambda^g_\text{eff}$ and $\Lambda^h_\text{eff}$ are respectively defined as $3/A^2_g$ and $3/A^2_h$ and they appear in the equations of motion as
\begin{align}
	&   3\dot{a}^2_g - \Lambda^g_\text{eff} a_g^4   =0 \\
	&  \dot{a_g}^2-2\ddot{a}_g a_g +\Lambda^g_\text{eff} a_g^4   =0 \, .
\end{align}
The above consistency relations mean that there exists a solution \textit{if and only if}
\begin{equation}
	\bar{\Lambda} a_g^2 a_h^2 + \Lambda_g a^4_g = \Lambda^g_\text{eff} a_g^4 \, ,
\end{equation}
where $\Lambda^g_\text{eff}$ is a constant. If we choose $\Lambda^g_\text{eff}=\Lambda^h_\text{eff} \equiv \Lambda_\text{eff}$ then by combining the two consistency relations we have,
\begin{equation}
	\Lambda_\text{eff} = \frac{\Lambda_g a_g^4 - \Lambda_h a_h^4}{a_g^4 - a_h^4} = \frac{\Lambda_g \nu_g - \Lambda_h \nu_h}{\nu_g - \nu_h} \, ,
\end{equation}
where the moir\'{e} relation is more clarified, with $\nu_{g,h}$ being spacetime volume elements of each universe in a global coordinate system.

The magnitudes of $\Lambda_h$, $\Lambda_g$ and $\bar{\Lambda}$, which directly appear in the action, are set by the scale of the theory to be either of order $M_P^2$ or zero. But $\Lambda_\text{eff}$ depends only on the constants of integration and therefore is arbitrarily chosen, which can be set to a very small value. For example, set $\Lambda^g_\text{eff} = \Lambda^h_\text{eff} \ll 1 $, $\bar{\Lambda}=-M_P^2$ and $\tau_h = \tau_g + \Delta \tau$. Then although at first $\Lambda_h$ and $\Lambda_g$ differ, they approach equality at long enough times.

We can also take the opposite path. Let us choose the $\Lambda^g_\text{eff} = \Lambda^h_\text{eff} = \bar{\Lambda}= M_P^2$ and $\tau_h = \tau_g + \Delta \tau$, with $\Delta \tau \gg 1$,
\begin{align}
	\Lambda_{g,h} = M_P^2 \left[1- \left(\frac{t + \tau_{g,h}}{t + \tau_{h,g}}\right)^2 \right] \, .
\end{align}
Then we get two universes that start way out of equilibrium with sizable cosmological constants, one negative and one positive, and gradually approach equilibrium at which their isolated cosmological constants vanish.

\acknowledgements  This work originated from a project supported by the \href{https://www.templeton.org}{Templeton Foundation}. This research was also supported by the Simons Foundation.

\bibliography{main}

\end{document}